\documentclass[aps,pra,twocolumn,showpacs,groupedaddress]{revtex4}
\usepackage{graphicx}
\usepackage{amsmath}
\usepackage{longtable}

\begin{document}

\title{The higher-order $C_{n}$ dispersion coefficients for hydrogen.} 
\author{J.Mitroy}  
\email{jxm107@rsphysse.anu.edu.au}
\author{M.W.J.Bromley}  
\email{mbromley@cdu.edu.au}
\affiliation{Faculty of Technology, Charles Darwin University, 
Darwin NT 0909, Australia}

\date{\today}

\begin{abstract}

The complete set of 2nd, 3rd and 4th order van der Waals 
coefficients, $C_n$ up to $n = 32$ for the H(1s)-H(1s) 
dimer are computed using pseudo-states to evaluate the 
appropriate sum rules.  A study of the convergence pattern 
for $n \le 16$ indicates that all the $C_{n \le 16}$ 
coefficients are accurate to 13 significant digits.  The 
relative size of the 4th-order $C^{(4)}_{n}$ to the 
2nd-order $C^{(2)}_{n}$ coefficients is seen to increase as 
$n$ increases and at $n=32$ the 4th-order term is actually
larger.    

\end{abstract}


\maketitle

\section{Introduction} 

The long-range interaction between two spherically 
symmetric atoms can be written in the general form 
\begin{equation}
V(R\to\infty) =  -V_6(R) - V_8(R) - V_{10}(R)    
 - V_{11}(R)  -   \ldots ,  
\label{vdw1}
\end{equation}
where the dispersion potential, $V_n(R)$ of order $n$ is
written  
\begin{equation}
V_n(R) = \frac{C_n}{R^n}   \ . 
\label{vdw2}
\end{equation}
The $C_n$ parameters are the van der Waals dispersion coefficients.  
The even ($n=6,8,\ldots$) dispersion coefficients are calculated 
using sum rules derived from 2nd-order perturbation theory and 
provide an attractive interaction.  The odd ($n=11,13,\ldots$) 
terms come from 3rd-order perturbation theory, and are repulsive 
\cite{ovsiannikov88a,yan99c}.  Contributions from 4th-order 
perturbation theory start at $n = 12$ \cite{bukta74a,ovsiannikov88a}.  

The dispersion interaction for the simplest system, namely the 
hydrogen dimer, is only known precisely to $n = 11$.   The
latest calculations by Yan and Dalgarno (YD) \cite{yan99c} reported
almost exact values for the 2nd-order even dispersion 
parameters up to $n = 16$.  They also gave almost exact results
for the 3rd-order coefficients, up to $n = 15$. However, it
is known that contributions from 4th-order perturbation 
theory start at $n = 12$ \cite{dalgarno67a,bukta74a,ovsiannikov88a}
so the dispersion potential computed from the YD $C_n$ is 
incomplete for $n > 11$.   
        
In this article, the complete set of dispersion parameters, 
from $C_{6}$ to $C_{16}$ are computed by using a large
pseudo-state expansion to evaluate the appropriate sum-rules.   
The contributions from 4th-order perturbation theory 
to $C_{12}$, $C_{14}$ and $C_{16}$ are explicitly included.

\section{Calculation of the dispersion parameters}

All the dispersion coefficients computed in this
paper were computed by first diagonalizing the
semi-empirical Hamiltonian in a large basis of 
Laguerre type orbitals (LTOs) defined by
\begin{equation}
\chi _\alpha (r)=N_\alpha r^{\ell+1} \exp (-\lambda _\alpha r)
L_{n_\alpha -\ell - 1}^{(2\ell +2)}(2\lambda _\alpha r) \ ,
\label{LTO}
\end{equation}
where the normalization constant is
                                                                            
\begin{equation}
N_\alpha =\sqrt{\frac{(2\lambda_\alpha)^{2\ell+3} (n_\alpha-\ell-1)!}
{(\ell+n_\alpha+1)!}} \ .
\end{equation}
The function $L_{n_\alpha-\ell-1}^{(2\ell +2)}(2 \lambda _\alpha r)$
is an associated Laguerre polynomial that can be defined in terms of
a confluent hypergeometric function \cite{abramowitz72a} as
\begin{eqnarray}
L_{n_\alpha-\ell-1}^{(2\ell +2)}(2\lambda _\alpha r)
&=& \frac{ (n_\alpha+\ell+1)!}{(n_\alpha-\ell-1)!(2\ell+2)!}  \nonumber \\ 
&\times &  M(-(n_\alpha-\ell-1),2\ell+2,2\lambda_\alpha r) \ .
\end{eqnarray}

All the matrix elements can be written in simple compact forms 
provided a common $\lambda_{\alpha}$ is adopted throughout
the calculation.  However, in the present work, the radial wave
functions were placed on a numerical grid and all matrix
elements computed by gaussian quadratures.  This was 
done for reasons on convenience as the diagonalisation 
could be done with an existing program used in previous 
calculations of the dispersion parameters and the structures
of positronic atoms \cite{bromley02a,mitroy03f}.  
This program can achieve close to machine precision in almost
all radial matrix computations.    
 
Once the Hamiltonian diagonalisation is complete, sum rules 
involving radial matrix elements were used to determine the
dispersion parameters.
The specific sum-rules used are those derived by 
Ovsiannikov, Guilyarovski and Lopatko (OGL) 
\cite{ovsiannikov88a}.  Their expressions are a bit 
simpler than those developed by other authors
\cite{bukta74a,yan99c}.  There were some omissions 
in their published equations \cite{ovsiannikov04a}, 
and a more thorough description of the sum-rules 
is presented here.   

\begin{table*}[th]
\caption[]{  \label{Chyd1}
The dispersion coefficients for the H-H dimer.  All
the results in the "Best Previous" column come from
\cite{yan99c} except that for $C^{(4)}_{12}$ which is 
taken from \cite{bukta74a}.  All values 
are in atomic units.
}
\vspace{0.1cm}
\begin{ruledtabular}
\begin{tabular}{lcccc}
Coefficient   & 10 LTOs & 15 LTOs & 20 LTOs  &  Best previous    \\
\hline
$C_6$ & 6.499 026 705 3305   &   6.499 026 705 4057  &  6.499 026 705 4059    &  6.499 026 705 4058   \\  
$C_8$ &  124.399 083 58236  &  124.399 083 58362  &  124.399 083 58362   &   124.399 083 58362    \\  
$C_{10}$ &  3285.828 414 9425 &   3285.828 414 9674  &   3285.828 414 9674   &     3285.828 414 9674    \\  
$C^{(2)}_{12}$ &  1.214 860 208 9619$\times 10^5$ &  1.214 860 208 9686$\times 10^5$ & 1.214 860 208 9686$\times 10^5$ &  1.214 860 208 9686$\times 10^5$ \\   
$C^{(2)}_{14}$ &  6.060 772 689 1671$\times 10^6$ &  6.060 772 689 1917$\times 10^6$ & 6.060 772 689 1917$\times 10^6$ &  6.060 772 689 1921$\times 10^6$ \\   
$C^{(2)}_{16}$ &  3.937 506 393 9865$\times 10^8$ &  3.937 506 393 9985$\times 10^8$ & 3.937 506 393 9985$\times 10^8$ &  3.937 506 393 9992$\times 10^8$  \\  
$C_{11}$ & -3474.898 037 8919   &  -3474.898 037 8822  & -3474.898 037 8822   & -3474.898 037 8822     \\  
$C_{13}$ &  -3.269 869 240 4549$\times 10^5$ &  -3.26 986 924 04407$\times 10^5$ & -3.26 986 924 04407$\times 10^5$ &  -3.26 986 924 04407$\times 10^5$ \\  
$C_{15}$ &  -2.839 558 063 3179$\times 10^7$ &  -2.839 558 063 2998$\times 10^7$ & -2.839 558 063 2997$\times 10^7$ &   -2.839 558 063 2998$\times 10^7$  \\ 
$C^{(4)}_{12}$ &  1241.587 803 8317  &    1241.587 803 8462  &   1241.587 803 8462  &  1241.588   \\  
$C^{(4)}_{14}$ &  3.009 633 558 9570$\times 10^5$  &  3.009 633 559 0035$\times 10^5$ &  3.096 633 559 0035 $\times 10^5$  &   \\  
$C^{(4)}_{16}$ &  4.745 455 287 4168$\times 10^7$  &  4.745 455 287 4083$\times 10^7$ &  4.745 455 287 4079$\times 10^7$    &   \\  
$C_{12}$ &  1.227 276 087 0002$\times 10^5$ &  1.227 276 087 0071$\times 10^5$ &    1.227 276 087 0071$\times 10^5$ &    1.227 27609$\times 10^5$ \footnotemark[1]  \\   
$C_{14}$ &  6.361 736 045 0628$\times 10^6$ &  6.361 736 045 0920$\times 10^6$ &  6.361 736 045 0921$\times 10^6$ &  \\   
$C_{16}$ & 4.412 051 922 7282$\times 10^8$ &  4.412 051 922 7393$\times 10^8$ &   4.412 051 922 7393$\times 10^8$ &   \\  
\end{tabular}
\end{ruledtabular}
\footnotetext[1]{ This entry adds to BM $C^{(4)}_{12}$ to the YD $C^{(4)}_{12}$.}
\end{table*}

\subsection{The 2nd-order terms}

The 2nd-order dispersion coefficients for the H-H system have 
been determined to high accuracy \cite{deal72a,thakkar88a,yan99c} 
even for high $n$.  The working expression adopted for computation 
is, 
\begin{eqnarray} 
C^{(2)}_{2\lambda+6} =  \sum_{\ell_1=1}^{\lambda+1} 
\frac { (2\lambda+4)! } { (2\ell_1+1)!(2\ell'_1+1)! }  \nonumber \\  
\times \sum_{i_1,i'_1}  
 \frac{ \langle 0,0 | r^{\ell_1} | i_1,l_1 \rangle^2 \langle 0,0 | r^{\ell'_1} | i'_1,\ell'_1 \rangle^2 }   
{ (E_{i_1} + E_{i'_1} - 2E_0)  }   
\label{C2N}
\end{eqnarray} 
where $\ell'_1 = \lambda + 2 - \ell_1$.  The state vector 
$| i_1,\ell_1 \rangle^2$ represents the radial part of the 
state $i_1$ with orbital angular momentum $\ell_1$ and
energy $E_{i_1}$.  The ground state energy is $E_0$. 
The sum rule 
\begin{equation} 
T^{(\ell)} = \sum_{i} \langle 0,0 | r^{\ell} | i,\ell \rangle^2 
                = \frac{(2\ell+2)!} {2^{(2\ell+1)}}    
\label{rsum1}
\end{equation} 
is a useful diagnostic check of the accuracy of the 
underlying discetization of the H-spectrum. 

\subsection{The 3rd-order terms; $C_{11}$ and $C_{13}$}

The dispersion coefficients, $C_{11}$ and $C_{13}$, arise from
3rd-order perturbation theory 
\cite{dalgarno67a,arrighini73a,bukta74a,ovsiannikov88a,yan99c}. 
Close to exact dispersion parameters for the 
H-H system have been published \cite{yan99c}. 

The general expression for the 3rd-order $C_{2\lambda+11}$ is \cite{ovsiannikov88a}  
\begin{eqnarray}
C_{2\lambda+11} & = & - \sum_{\ell_1 k_1 \ell_2} \sum_{\ell'_1 k'_1 \ell'_2} \sum_{i_1 i'_1 i_2 i'_2}  \nonumber \\   
 & \times &  \frac{G(\lambda,\ell_1,\ell'_1,\ell_2,\ell'_2,k_1,k'_1)} 
  {(E_{i_1 i'_1}-2E_0)(E_{i_2 i'_2}-2E_0)} \nonumber \\  
 & \times &  \langle 0,0 | r^{\ell_1} |i_1,\ell_1\rangle     
 \langle i_1,\ell_1 | r^{k_1} | i_2,\ell_2 \rangle  
 \nonumber \\  
  & \times & 
 \langle i_2,\ell_2 | r^{\ell_2}  | 0,0 \rangle     
 \langle 0,0 | r^{\ell'_1}  | i'_1,\ell'_1 \rangle   
 \nonumber \\ 
 &\times& 
 \langle i'_1,\ell'_1 | r^{k'_1}  |i'_2,\ell'_2\rangle     
 \langle i'_2,\ell'_2 |  r^{\ell'_2}  | 0,0\rangle      
\end{eqnarray}
with the notation $E_{i_1 i'_1} = E_{i_1} + E_{i'_1}$ being used
in the energy denominator.  The parameter $\lambda$ is defined  
\begin{equation} 
2\lambda +8 = \ell_1 + k_1 + \ell_2 + \ell_1' + k'_1 + \ell'_2 
\end{equation} 
and all of the angular momentum indices are greater than zero. Defining  
$J = (\ell_1 + k_1 + \ell_2)/2$ and $J' = (\ell'_1 + k'_1 + \ell'_2)/2$, 
the coefficient $G$ is defined as     
\begin{eqnarray}
G(\lambda,\ell_1,\ell'_1,\ell_2,\ell'_2,k_1,k'_1) =   (\lambda+4)! A(J,\ell_1,k_1,\ell_2)  \nonumber \\  
\times A(J',\ell'_1,k'_1,\ell'_2) B(\lambda,\ell_1,\ell'_1)  B(\lambda,k_1,k'_1) B(\lambda,\ell_2,\ell'_2) \nonumber \\    
\end{eqnarray}
where 
\begin{equation}
B(\lambda,\ell_1,\ell_2) = \frac{[2(\lambda+4-\ell_1-\ell_2)]!} {(\lambda+4-\ell_1-\ell_2)!} \ ,   
\end{equation}
and
\begin{equation}
A(J,\ell_1,k_1,\ell_2) = \frac{J!} {(2J+1)!(J-\ell_1)!(J-k_1)!(J-\ell_2)!} .  
\end{equation}

\subsection{The 4th-order contributions to $C_{12}$ and $C_{14}$}

As far as we know, there have only been two explicit calculations 
of the 4th-order contribution to $C_{12}$.  Bukta and Meath   
\cite{bukta74a} gave estimates of $C^{(2)}_{12}$ and $C^{(4)}_{12}$  
for the hydrogen dimer.  Ovsiannikov {\em et al} 
\cite{ovsiannikov88a} developed a general and compact expression 
for the evaluation of $C^{(4)}_{n}$, and in addition they 
reported values of $C^{(4)}_{12}$ for all possible combinations 
of hydrogen and the alkali atoms.  Rectifying some omissions 
in their published equations \cite{ovsiannikov04a}, one writes   

\begin{equation}
C^{(4)}_{2\lambda + 12} = b_{2\lambda+12} - \sum^{\lambda}_{\lambda_1=0} C^{(2)}_{2\lambda_1+6} a_{2\lambda_2+6} \ ,  
\end{equation}
where
\begin{equation}
\lambda = \lambda_1 + \lambda_2 \ .  
\end{equation}
The factor $a_{2\lambda_2+6}$ is  
\begin{eqnarray}
a_{2\lambda_2+6} =  \sum_{\ell_1=1}^{\lambda_2+1} 
\frac { (2\lambda_2+4)! } { (2\ell_1+1)!(2\ell'_1+1)! }  \nonumber \\  
\times \sum_{i_1,i'_1}  
 \frac{ \langle 0,0 | r^{\ell_1} | i_1,l_1 \rangle^2 \langle 0,0 | r^{\ell'_1} | i'_1,\ell'_1 \rangle^2 }   
{ (E_{i_1} + E_{i'_1} - 2E_0)^2  }   
\label{A2N}
\end{eqnarray} 
where   
\begin{equation}
\ell_1 + \ell'_1 = \lambda + 2 \ .   
\end{equation}
The expression for $a_{2\lambda+6}$ is practically the same 
as eq.~(\ref{C2N}) for $C^{(2)}_{2\lambda+6}$; the only
difference being an extra factor in the energy denominator 
(compare with eq.~(10) of \cite{ovsiannikov88a}). 

The factor $b_{2\lambda+12}$ is more complicated and defined as  
\begin{eqnarray}
b_{2\lambda+12} &=& \sum_{\ell_1,\ell_2,\ell_3} \sum_{\ell'_1,\ell'_2,\ell'_3}  
\sum_{k_1,k_2,k'_1,k'_2} \sum_{K} \sum_{i_1 i'_1 i_2 i'_2 i_3 i'_3 }  \nonumber \\  
& \times & \left[ \frac{{\hat L}_1! {\hat K}_1! {\hat K}_2! {\hat L}_3! }
 {2\ell_1!2\ell'_1!2k_1!2k'_1!2k_2!2k'_2!2\ell_3!2\ell'_3!} \right]^{1/2}  \nonumber \\  
&\times& \langle \ell_1 0 k_1 0 | \ell_2 0 \rangle  
\langle \ell'_1 0 k'_1 0 | \ell'_2 0 \rangle  \nonumber \\  
&\times& \langle k_2 0 \ell_3 0 | \ell_2 0 \rangle  
\langle k'_2 0 \ell'_3 0 | \ell'_2 0 \rangle  \nonumber \\  
&\times& \langle L_1 0 K_1  0 | K 0 \rangle  
\langle K_2 0 L_3 0 | K 0 \rangle  \nonumber \\  
& \times & \left\{  \begin{array}{ccc}  
\ell_1 & \ell'_1 & L_1 \\
k_1 & k'_1 & K_1 \\  
\ell_2 & \ell'_2 & K   
\end{array} \right\}  
\left\{  \begin{array}{ccc}  
k_2 & k'_2 & K_2 \\
\ell_3 & \ell'_3 & L_3 \\  
\ell_2 & \ell'_2 & K    
\end{array} \right\}  \nonumber \\  
&\times& 
\frac{1}{(E_{i_1 i_1'}-2E_0)(E_{i_2 i'_2}-2E_0)(E_{i_3 i'_3}-2E_0)} \nonumber \\ 
&\times &  \langle 0,0 | r^{\ell_1} | i_1,\ell_1  \rangle    
\langle i_1,\ell_1 | r^{k_1} | i_2,\ell_2 \rangle  \nonumber \\   
& \times & \langle i_2,\ell_2 | r^{k_2} | i_3,\ell_3 \rangle     
\langle i_3,\ell_3 | r^{\ell_3} | 0,0 \rangle  \nonumber \\   
&\times &  \langle 0,0 | r^{\ell'_1} | i'_1,\ell'_1  \rangle    
\langle i'_1,\ell'_1 | r^{k'_1} | i'_2,\ell'_2 \rangle  \nonumber \\   
& \times & \langle i'_2,\ell'_2 | r^{k'_2} | i'_3,\ell'_3 \rangle     
\langle i'_3,\ell'_3 | r^{\ell'_3} | 0,0 \rangle      
\label{b4n} 
\end{eqnarray}
where 
$L_1 = \ell_1 + \ell'_1$, $L_3 = \ell_3 + \ell'_3$, 
$K_1 = k_1 + k'_1$ and $K_2 = k_2 + k'_2$. We use 
${\hat L} = (2L+1)$.    
The sums are constrained by the condition  
\begin{equation}
L_1 + K_1 + K_2 + L_3 =  2\lambda + 8  \ .   
\end{equation}
While $\ell_1$, $\ell'_1$, $\ell_3$, $\ell'_3$ must
be greater than 0, it is possible for $\ell_2$ and 
$\ell'_2$ to be equal to 0.   None of  $k_1$, $k'_1$, 
$k_2$ or $k'_2$ can be zero.  Since $\ell_2$ and 
$\ell'_2$ can both be equal to zero, the possibility 
of $i_2,i'_2$ both occupying the ground state must 
be explicitly excluded from the summation.   

\section{Results of the calculation}

The results of the calculations for the complete set
of dispersion coefficients up to $C_{16}$ are given in Table 
\ref{Chyd1}.    The parameters are given for basis sets with 10, 
15 and 20 basis functions per angular momentum respectively.
The exponent in the LTO was chosen to be $\lambda = 1.0$ 
for all angular momenta.  This choice resulted in much
faster convergence of the dispersion parameters than that
observed by Yan and Dalgarno in their calculations 
of the 3rd-order dispersion coefficients.  Table \ref{Chyd1} 
also gives results reported by YD and a single calculation of 
$C^{(4)}_{12}$ by Bukta and Meath (BM) \cite{bukta74a}.

\subsection{The 2nd-order terms}

The calculations of $C^{(2)}_n$ do not give new information
and Yan and Dalgarno \cite{yan99c} have given values which
are converged to better than 15 significant figures.  The 
present calculations with the $N=20$ basis are identical 
to 13 significant figures.  The small differences in the
last digit for some of coefficients arise from minor inaccuracies
with the radial matrix elements.  Hence we conclude that 
the present calculations are numerically reliable and that 
the pseudo-state representation of the H-spectrum is
close to converged.   

Besides the dispersion coefficients, the sum-rule, eq.~(\ref{rsum1}) 
was evaluated and seen to be correct to 12 significant digits 
for all polarities relevant to the evaluation of $C_6$ to
$C^{(2)}_{16}$.

\begin{table*}[th]
\caption[]{  \label{Chyd2}
The $n \ge 16$ dispersion coefficients for the H-H dimer.  
All values are in atomic units.
}
\vspace{0.1cm}
\begin{ruledtabular}
\begin{tabular}{lcccc}
$n$   & $C^{(2)}_{n}$ &    $C^{(4)}_{n}$  & $C^{(2)}_n+C^{(4)}_{n}$ & $C^{(3)}_{n}$  \\ 
\hline   
17    &  &    &   & 2.726 099 889$\times10^{9}$   \\   
18    & 3.234 218 716$\times10^{10}$   &  7.009 061 179$\times10^{9}$  & 3.935 124 834$\times10^{10}$   &  \\   
19    &  &    &   &  3.020 900 833$\times10^{11}$   \\   
20   &  3.278 573 440$\times10^{12}$    &  1.083 922 188$\times10^{12}$  & 4.362 495 628$\times10^{12}$   &  \\   
21    &  &    &   &   3.900 227 980$\times10^{13}$    \\   
22    & 4.021 082 848$\times10^{14}$   &  1.832 218 347$\times10^{14}$  & 5.853 301 195$\times10^{14}$  &   \\   
23    &  &   &   &    5.856 636 712$\times10^{15}$   \\   
24    &  5.868 996 335$\times10^{16}$   &  3.444 924 821$\times10^{16}$  & 9.313 921 156$\times10^{16}$  &  \\   
25    &  &    &   &    1.017 059 252$\times10^{18}$   \\   
26    & 1.005 294 993$\times10^{19}$  & 7.249 737 286$\times10^{18}$  &  1.730 268 722$\times10^{19}$   &      \\   
27    &  &    &   &     2.028 440 001$\times10^{20}$   \\   
28    &  1.996 944 941$\times10^{21}$   &  1.709 243 726$\times10^{21}$ &  3.706 188 667$\times10^{21}$  &       \\   
29    &  &    &   &    4.613 037 362$\times10^{22}$   \\   
30    &  4.553 288 866$\times10^{23}$  & 4.507 006 859$\times10^{23}$  &  9.060 295 725$\times10^{23}$  &      \\   
31   &  &    &   &    1.188 007 684$\times10^{25}$   \\   
32    &  1.181 107 088$\times10^{26}$  &  1.325 398 446$\times10^{26}$  & 2.506 505 534$\times10^{26}$  &      \\   
\end{tabular}
\end{ruledtabular}
\end{table*}

\subsection{The 3rd-order terms}

Since the 3rd-order terms, $C^{(3)}_n$ have already been 
given by YD, these calculations merely serve as a test of 
our numerical procedures. Once again, calculations with the
20 LTO basis agree with the YD results to 14 significant 
figures.  It is worth noting the present results 
required fewer terms than YD to achieve convergence. 
YD made the choice $\lambda_{\alpha} = 1/(\ell+1)$ in 
eq.~(\ref{LTO}) and did not achieve convergence to the
14th digit place until the dimension of the LTO expansion
was 50.  The present basis with $\lambda_{\alpha} = 1.0$ 
achieves the same level of convergence with 20 LTOs. 
 
\subsection{The 4th-order terms}

The only previous explicit calculation of a 4th-order term was
that made by Bukta and Meath (BM) \cite{bukta74a}, and the only
parameter  given was $C^{(4)}_{12}$.  The OGL 
\cite{ovsiannikov88a} estimate of $C^{(4)}_{12}$,
1.220$\times 10^5$ au was made using an approximation to 
the Greens function and so perfect agreement is not
expected.   However, the present calculation agrees with
BM calculation of $C^{(4)}_{12}$ to all digits quoted, 
namely seven.      

The number of terms in the sum, eq.~(\ref{b4n}) increases rapidly 
as $n$ increases.  There are 4 terms for $C^{(4)}_{12}$,  
there are 64 terms for $C^{(4)}_{14}$, and finally there are 
460 terms for $C^{(4)}_{16}$.  

The dominant contribution to $C^{(4)}_{n}$ comes
from $b_{2\lambda+12}$ with 96$\%$ of $C^{(4)}_{12}$ 
coming from $b_{12}$.  The tendency for $b_{2\lambda+12}$ 
to be the dominant term in $C^{(4)}_{n}$ becomes more
accentuated as $n$ increases and the $b_{16}$ 
term gives an estimate to $C^{(4)}_{16}$ that is 
correct to 0.1$\%$.

One feature of of Table \ref{Chyd1} concerns the relative
size of $C^{(4)}_{n}$ to $C^{(2)}_{n}$.  For $n = 12$, the
$C^{(4)}_{n}:C^{(2)}_{n}$ ratio is 1.02$\%$.  However, as $n$
gets larger, the ratio also gets larger.  For $n = 14$ 
the ratio is 4.97$\%$, while for $n = 16$ the ratio is 
12.1$\%$.   

\subsection{The dispersion coefficients for $n \ge 17$ }

Higher order contributions than 4th-order begin at 
$n = 17$.  There is a 5th-order contribution to $C_{17}$
and a 6th-order contribution to $C_{18}$ \cite{ovsiannikov88a}.   
Estimates of $C^{(2)}_{n}$ for $n \ge 17$ have been made 
by a variety of authors 
\cite{bell65a,bell66a,koide81a,thakkar88a,ovsiannikov88a}.  
However, the only estimate of the 3rd and 4th-order
terms with $n \ge 17$ are those of OGL \cite{ovsiannikov88a}.  
By explicit calculation they obtained 
$C^{(3)}_{17} = -2.739\times10^9$ au which agrees with
the present more extensive calculation to within 1$\%$.  
Making an approximation to the greens function  
they estimated $C^{(4)}_{18} = 3.3\times10^9$ au, which
is about half the size of the present value.

The dispersion parameters up to $C_{32}$ from the present  
calculation are tabulated in Table \ref{Chyd2}.  The reason
for taking the calculations so far rests in the relative
size of the $C^{(4)}_{n}$ and $C^{(2)}_{n}$ terms.  It 
was noticed that the $C^{(4)}_{n}$:$C^{(2)}_{n}$ ratio
got larger as $n$ increased.  So the calculations were
extended to $C_{32}$ in order to demonstrate explicitly 
that the $C^{(4)}_{n}$:$C^{(2)}_{n}$ ratio can actually  
become larger than 1.0.  

The precision of the entries in Table \ref{Chyd2} is not as 
high as those in Table \ref{Chyd1}.  The calculations of 
$C^{(4)}_{n}$ did become more time consuming as $n$ increased.  
There were 922,064 different $(\ell_1,k_1,\ell_2,\ldots)$ 
combinations by the time $n=32$ was reached.   Also the 
number of radial integrals in eq.~(\ref{b4n}) increases 
as $N^6$ where $N$ is the numbers of LTOs for any given 
$\ell$.  So the $N=20$ calculation is 64 times more intensive 
than the $N = 10$ calculation.        

The $C^{(2)}_{n}$ and $C^{(3)}_{n}$ entries in Table \ref{Chyd2} 
were taken from the calculation with 15 LTOs. The $C^{(4)}_{n}$ entries 
were taken from a $N = 15$ calculation up to $n = 20$, thereafter 
the $N = 10$ basis was used.  The values of $C^{(2)}_{n}$ agree
with those of Thakkar \cite{thakkar88a} for all ten digits
given in Table \ref{Chyd2}.  Comparisons between $N = 10$ and
$N = 15$ calculations for $C^{(3)}_{n}$ suggest that the
convergence is slower as $n$ increases and that $C^{(3)}_{31}$ 
is reliable to about 6 digits.  A similar level of accuracy 
can be expected for $C^{(4)}_{n}$ and a comparison between
the $N=10$ and $N=15$ values for $C^{(4)}_{20}$ gives agreement 
for the first 9 digits.  

\section{Conclusions}

The higher $n$ dispersion parameters from $C_{11}$, 
through to $C_{16}$ have been computed to an accuracy of 
13 significant figures for the H-H dimer.  Since the 4th-order     
contributions were included for $C_{12}$, $C_{14}$ and $C_{16}$
the adiabatic dispersion interaction can now be regarded as
complete up to terms of order $R^{-16}$.   

The time taken to evaluate the dispersion coefficients 
was not excessive.  For example, a calculation using 20 LTOs 
took about 17 minutes to determine all terms up $C_{16}$ on 
a 850 MHz CPU.  Hence the pseudo-state method adopted here, 
and in other similar works (e.g. \cite{bukta74a,yan99c}), 
could be used to make explicit calculations of the 
5th-order correction to $C_{17}$ and even the 6th-order 
correction to $C_{18}$ \cite{ovsiannikov88a}.  Therefore, it 
is certainly possible
with existing technology to determine the complete dispersion 
interaction for the H-H interaction for all terms up to 
and including $C_{22}$. 

\section{Acknowledgments}

The authors would like to thank Mr J C Nou and Mr C Hoffman  
of CDU for workstation support and Professor V D Ovsiannikov 
of Voronezh University for helpful communications about the
exact form of the 4th-order matrix element and for pointing
out some faults with this manuscript.

\end{document}